\documentclass[preprint]{aastex}
\usepackage{bm,amsmath}
\shorttitle{Constraining Electron Spectra in Mrk~421}
\shortauthors{Ushio et al.}

\begin{document}

\title{A Novel Approach in Constraining Electron Spectra in Blazar Jets:\\ The Case of Markarian~421}

\author{Masayoshi Ushio\altaffilmark{1,\,2}, {\L}ukasz Stawarz\altaffilmark{1,\,3}, Tadayuki Takahashi\altaffilmark{1,\,2}, David Paneque\altaffilmark{4}, Grzegorz Madejski\altaffilmark{4}, Masaaki Hayashida\altaffilmark{4}, Jun Kataoka\altaffilmark{5}, Yasuyuki T. Tanaka\altaffilmark{1}, Takaaki Tanaka\altaffilmark{4}, and Micha{\l} Ostrowski\altaffilmark{3}}

\email{E-mail:ushio@astro.isas.jaxa.jp}

\altaffiltext{1}{Department of High Energy Astrophysics, Institute of Space and Astronautical Science (ISAS), Japan Aerospace Exploration Agency (JAXA), 3-1-1 Yoshinodai, Sagamihara, 229-8510 Japan}
\altaffiltext{2}{Department of Physics, University of Tokyo, Hongo 7-3-1, Bunkyo, 113-0033 Japan}
\altaffiltext{3}{Astronomical Observatory, Jagiellonian University, ul. Orla 171, Krak\'ow 30-244, Poland}
\altaffiltext{4}{Stanford Linear Accelerator Center/KIPAC, 2575 Sand Hill Road, Menlo Park, CA 94025, USA}
\altaffiltext{5}{Research Institute for Science and Engineering, Waseda University, 3-4-1 Okubo, Shinjuku, Tokyo, 169-8555 Japan}

\begin{abstract}
We report results from the observations of the well studied TeV blazar Mrk~421 with the {\it Swift} and the {\it Suzaku} satellites in December 2008. During the observation, Mrk~421 was found in a relatively low activity state, with the corresponding $2-10$\,keV flux of $3\times 10^{-10}$\,erg\,s$^{-1}$\,cm$^{-2}$.  For the purpose of robust constraining the UV--to--X-ray emission continuum we selected only the data corresponding to truly simultaneous time intervals between {\it Swift} and {\it Suzaku}, allowing us to obtain a good-quality, broad-band spectrum despite a modest length ($0.6$\,ksec) exposure. We analyzed the spectrum with the parametric forward-fitting {\tt SYNCHROTRON} model implemented in {\tt XSPEC} assuming two different representations of the underlying electron energy distribution, both well motivated by the current particle acceleration models: a power-law distribution above the minimum energy $\gamma_{\rm min}$ with an exponential cutoff at the maximum energy $\gamma_{\rm max}$, and a modified ultra-relativistic Maxwellian with an equilibrium energy $\gamma_{\rm eq}$. We found that the latter implies unlikely physical conditions within the blazar zone of Mrk~421. On the other hand, the exponentially moderated power-law electron distribution gives two possible sets of the model parameters: (i) flat spectrum $dN'_e/d\gamma \propto \gamma^{-1.91}$ with low minimum electron energy $\gamma_{\rm min}<10^3$, and (ii) steep spectrum $\propto \gamma^{-2.77}$ with high minimum electron energy $\gamma_{\rm min}\simeq 2\times10^4$. We discuss different interpretations of both possibilities in the context of a diffusive acceleration of electrons at relativistic, sub- or superluminal shocks. We also comment on how exactly the $\gamma$-ray data can be used to discriminate between the proposed different scenarios.
\end{abstract}


\section{Introduction} 
\label{sect:introduction}

Blazars constitute a sub-class of radio-loud, jet-dominated 
Active Galactic Nuclei (AGN) characterized by a broad-band 
non-thermal emission detected from radio to high energy 
$\gamma$-ray band, strong variability on time scales ranging 
from hours up to years, two distinct (low- and high-energy) 
spectral humps in the $\nu$--$\nu F_{\nu}$ representation, 
and also by a substantial polarization in the radio and optical 
photon energy ranges. The low-energy spectral component of blazar 
sources, peaking at IR--to--X-ray frequencies, is widely believed 
to be due to a Doppler-boosted synchrotron emission of 
ultra-relativistic electrons accelerated in the innermost 
parts of relativistic and oriented close to our line of sight jets 
containing magnetized plasma \citep{urry95}. The high-energy spectral 
component of low-power blazars of the BL Lacertae type (hereafter 
BL Lacs) extends up to a few or even tens of TeV, and is most 
commonly considered as originating via inverse-Compton scattering 
by ultra-relativistic electrons of synchrotron photons produced 
by the same electrons (``synchrotron self-Compton process,'' 
hereafter SSC). This $\gamma$-ray emission, which only recently can 
be studied in detail with the new generation of ground-based Imaging 
Atmospheric Cherenkov Telescopes (IACTs) and {\it Fermi} satellite, 
exhibits particularly dramatic variability, observed on the timescales 
as short as minutes \citep{aha07,alb07}. This indicates extremely 
efficient and rapid particle acceleration processes taking place in 
compact emission regions. Our understanding of such acceleration 
processes is, on the other hand, still insufficient, despite a 
significant theoretical progress made in this field during the last years.

Markarian~421 (Mrk~421; $z=0.03$), the first established extragalactic 
TeV source \citep{punch92}, is one of the most famous and nearby 
``high frequency peaked'' BL Lacs. Since its discovery, it has 
been routinely observed by several IACTs at very high energy 
$\gamma$-rays and, more importantly, it has been targeted by 
many intensive multi-wavelength campaigns (which, in the past, were 
limited in their temporal and spectral coverage, and notably, often were 
triggered by a particularly high activity state of the source).  
These resulted in the discovery of a significant correlation 
between the hard X-ray and TeV fluxes, thus supporting 
the SSC origin of its high-energy emission 
\citep[e.g.][]{inoue96,tad96,fossati08}. In a framework of such 
an interpretation, the energy distribution of the radiating electrons 
is required to be of a broad, non-thermal form, and the power-law 
spectral shape is the most commonly invoked approximation for it. 
The main justification for this approximation is that 
the diffusive shock acceleration, the ``1st-order Fermi process'' --- 
known to generate power-law spectra of ultra-relativistic particles --- 
is likely to play a major role in accelerating jet electrons 
\citep[e.g.][]{mas97,tad00,spada01}. Still, as pointed out by 
several authors, alternative forms of the electron energy distribution 
in blazar jets, namely ``ultra-relativistic Maxwellian'' expected 
to be produced in the case of the efficient turbulent (``2nd-order 
Fermi'') acceleration, may also account for bulk of the observational 
findings \citep[e.g.][]{sau04,kat06b,gie07,ushio2009}.  In order to 
determine which acceleration process is likely to be dominant, it is
necessary to reconstruct the underlying electron spectrum from the observed 
emission spectra. The X-ray observations are crucial in this respect.  

The {\it Suzaku} X-ray observatory \citep{mitsuda06} has the capability 
of high quality observations of astrophysical objects from $0.5$ up 
to $300$\,keV, owing to the excellent sensitivity of its Hard X-ray Detector 
\citep[HXD;][]{takahashi_hxd06}. As such, it is the most suitable 
instrument for analyzing the synchrotron continua of high-energy 
peaked BL Lacs. Recently, \citet{ushio2009} demonstrated the excellent 
quality of the {\it Suzaku} data by constructing high-photon-statistics 
time-resolved $0.6-60$\,keV spectra of Mrk~421 for exposure times 
as short as a few ksec. These were analyzed by the parametric 
forward-fitting {\tt SYNCHROTRON} model (rather than being fitted 
by some assumed spectral shape of the synchrotron continuum), to 
investigate more directly the electron energy distribution.  
Here, we proceed further in this direction, by fitting the broad-band 
spectrum of Mrk~421 with different forms of the electron distribution 
discussed in the literature. In addition to the new {\it Suzaku} data 
(collected during December 2008), however, we also utilize the 
simultaneous observations by the {\it Swift} satellite \citep{gehrels2004}, 
and in particular the Ultra-Violet/Optical Telescope 
\citep[UVOT;][]{roming05} onboard {\it Swift}. \S\,\ref{sect:observation} 
reports in more detail on the analyzed datasets. In \S\,\ref{sect:analysis}, 
we present the results of the applied spectral analysis. The implications 
of our results towards the understanding of blazar sources are 
discussed and summarized in \S\,\ref{sect:discussion} and 
\S\,\ref{sect:conclusions}. Throughout this analysis, we use the 
{\tt XSPEC} version 12.5.0ac and the {\tt CALDB} of version 2009.04.03.

\section{Observations and Data Reduction} 
\label{sect:observation}

\subsection{{\it Suzaku}}
\label{sect:suzaku}

{\it Suzaku} observations were performed with the X-ray Imaging Spectrometer 
\citep[XIS;][]{koyama06} in the $0.3-12$\,keV band, and also with the 
Hard X-ray Detector \citep[HXD;][]{takahashi_hxd06,kokubun06} in the 
$10-300$\,keV band. The XIS cameras, located at the foci of the X-ray 
telescopes \citep[XRTs;][]{serlemit06}, include one back-illuminated 
CCD camera (BI) and two front-illuminated CCD cameras (FIs). The HXD 
is a non-imaging detector system which consists of silicon PIN 
diodes capable of observing in the $10-70$\,keV band and the GSO 
crystal scintillators in the $40-300$\,keV band. In the following analysis, 
we do not use the GSO data, because Mrk~421 was not sufficiently 
bright above $40$\,keV during the performed observations to be 
significantly detected by the GSO.  

We observed Mrk~421 with {\it Suzaku} for almost three days from 
MJD 54803.76 to MJD 54805.89 (seqnum=703020010). All of the XISs 
are operated with 1/4 window option in order to reduce possible pile-up 
effects. Both XIS and HXD data are reprocessed from ``uncleaned event 
files'' with the {\tt CALDB} of 2009-04-07 which corresponds to 
{\tt rev2.2} of the official pipeline process. The ``unscreened event 
files'' of XISs are reduced with the standard selection criteria requiring, 
in addition, the rigidity larger than $6$\,GV/$c$. The XIS events are next 
extracted from a circular region with a radius of $3'$ centered on the 
image peak. The background region is extracted from the source free region 
having a similar area. Moreover, we extract the image center of BI 
sensor with a radius of $0.2'$ because it weakly suffers from pile-up 
judging from its count rate. In our spectral fitting, we derive the 
response matrices (RMF) and the effective area files for the XRTs (ARF) 
for each XIS sensor by using {\tt xisrmfgen} and {\tt xissimarfgen}, 
respectively \citep{ishisaki06}. Since the two FI CCDs have almost 
identical performance, we co-add their data and the corresponding 
RMF and ARF. For the HXD data, we exclude events during South 
Atlantic Anomaly (SAA) passages, Earth occultation and those 
with cutoff rigidity less than $6$\,GV/c. 

\subsection{{\it Swift}}
\label{sect:swift}

The three-day-long {\it Suzaku} observation of Mrk~421 is well complemented 
by two {\it Swift} observations, involving nine short exposures 
($\sim 200$\,s each; seqnum=00030352102 and 00030352103). The {\it Swift} 
satellite carries three sets of instruments: the Burst Alert Telescope 
\citep[BAT; $15-150$\,keV;][]{gehrels2004}, the X-ray Telescope \citep[XRT; 
$0.3-10$\,keV;][]{burrow05}, and the Ultra-Violet/Optical Telescope 
\citep[UVOT; $170-650$\,nm;][]{roming05}. Hereafter, we only analyze 
the XRT and the UVOT data. All the data are taken from the NASA/{\it Swift} 
database (\texttt{http://heasarc.gsfc.nasa.gov/FTP/swift/}) and 
are processed at the {\it Swift} Data Center. 

The {\it Swift} XRT data were all taken in the Windowed Timing mode, with 
the central $8$\,arcmin field of view covered, and one dimensional imaging 
preserved. Therefore, we extract the rectangular region of 
$0.50\times 3.0$\,arcsec$^2$ centered on the peak position of the source. 
The background is selected in the source free region with rectangular 
shape. Photons with grade $0-2$ are selected and rebinned so that each 
bin has sufficient photon statistics. The auxiliary response file is created 
by the task {\tt xrtmkarf} and the standard response file 
{\tt swxwt0to2s6\_20070901v011.rmf}. For UVOT observations, we utilize 
the three ultraviolet filters in the imaging mode, with the effective 
central wavelength of $260.0$\,nm, $224.6$\,nm and $192.8$\,nm for UVW1, 
UVM2 and UVW2 filters, respectively. The source aperture sizes are all 
chosen to correspond to those used to determined the UVOT zero points, 
namely $5^{\prime\prime}$ for the ultraviolet filters. A $15^{\prime\prime}$ 
background region is extracted from the ``ghost wing'' of nearby bright 
source 51~UMa. A comparison of results using several different background 
regions suggests that the corresponding uncertainty is at most $0.01$\,mag 
\citep{poole08}, which is smaller than the systematic uncertainty indicated 
by the UVOT team. Next, we calculate the photometric spectra so that they 
can be fitted in {\tt XSPEC} by the task {\tt uvot2pha} (v1.5) using 
the response matrices {\tt swu[{\tt w1},{\tt m2},{\tt w2}]\_20041120v104.rsp}. 
Note that the columns ``E\_MIN'' and ``E\_MAX'' of these response are 
manually set to the FWHM of the filter in order to improve their appearance. 

\section{Analysis} 
\label{sect:analysis}

\subsection{The {\tt SYNCHROTRON} Model}
\label{sect:model}

The broad bandpass and high-sensitivity of the {\it Suzaku} satellite 
enables spectral fits for the observed X-ray emission of celestial 
sources with synchrotron spectra originating from a given (assumed) 
shape of the electron energy distribution (EED), rather than by the 
assumed spectral form of the non-thermal emission continuum 
\citep{ttanaka08,ushio2009}. We note that in the past the analogous 
``parametric forward-fitting'' technique has been applied in analyzing 
X-ray emission of solar flares, believed to be produced by bremsstrahlung 
of non-thermal electrons within the magnetic loops above the solar surface 
\citep[see][and references therein]{park97,petrosian10}. In this paper, 
we continue the studies of the X-ray emission of Mrk~421 initiated in 
\citet{ushio2009}, including the ultraviolet observations by 
UVOT onboard {\it Swift} satellite in addition to the analyzed 
{\it Suzaku} dataset. Below we briefly describe the {\tt SYNCHROTRON} 
model serving this function, as implemented in {\tt XSPEC}.

In our {\tt SYNCHROTRON} model, we assume two different forms for 
the EED, both well motivated by the particle acceleration models 
discussed in the literature (see next section), namely a power-law 
(PL) form above the minimum energy $\gamma_{\rm min}$ with an exponential 
cutoff at the maximum energy $\gamma_{\rm max}$, and a modified 
ultra-relativistic Maxwellian (UM) with the equilibrium 
energy $\gamma_{\rm eq}$:
\begin{equation}
\frac{dN'_e}{d\gamma} = \left\{
\begin{array}{llll}
	N'_0 \, \gamma^{-s} \, \exp\!\left[ - \left(\frac{\gamma}{\gamma_{\rm max}}\right)\right] & {\rm for} & 1 \leq \gamma_{\rm min} \leq \gamma & {\rm (PL)} \vspace{0.1in} \\
	N'_0 \, \gamma^2 \, \exp\!\left[ -\frac{1}{a} \left(\frac{\gamma}{\gamma_{\rm eq}}\right)^a \right] & {\rm for} & 1 \leq \gamma & {\rm (UM)}
\end{array}
\right. \, ,
\label{eq:dist}
\end{equation}
where $\gamma \equiv E'_e/m_e c^2$ is the electron Lorentz factor, 
$N'_{0}$ is the normalization of the EED, $s$ is the electron energy 
index, and $a$ represents the modulation of the exponential cut-off. 
Here and below the primes denote the quantities measured in the 
emitting plasma rest frame, except of electron Lorentz factor $\gamma$, 
which is always measured in the this frame. The considered model PL is 
thus characterized by three free parameters ($s$, $\gamma_{\rm min}$, 
and $\gamma_{\rm max}$), while the model UM by two free parameters 
only ($a$ and $\gamma_{\rm eq}$; since with the extremely flat 
low-energy spectrum $\propto \gamma^2$ the exact location of 
$\gamma_{\rm min}$ does not affect the modeled synchrotron continuum). 
For these, we consider the following ``expected'' ranges: 
$1.5 \leq s \leq 3.5$, $1 \leq \gamma_{\rm min} < \gamma_{\rm max}$, 
$10^{4.5} \leq \gamma_{\rm max} \leq 10^{6.5}$, $0.1 \leq a \leq 2$, 
and $10^{1.0} \leq \gamma_{\rm eq} \leq 10^{5.0}$. We note that this 
is the first time when the two aforementioned representations of 
the EED are tested against a single, broad-band dataset.

Having specified the form of the EED with the assumed 
isotropic pitch-angle distribution, the jet comoving synchrotron 
emissivity $j'_{\rm syn}(\epsilon')$ is described by: 
\begin{equation}
	j'_{\rm syn}(\epsilon') \propto B' 
	\int \frac{d N'_e}{d\gamma} \,\,\, F\!\!\left(\frac{\epsilon'}{\epsilon'_c} \right) \,\, d\gamma \, ,
\end{equation}
where the function $F(x)$ is defined as
\begin{equation}
	F(x) \equiv x \int_x^\infty K_{5/3}(\xi) \,\, d\xi \, ,
\end{equation}
and where $K_{5/3}(\xi)$ is the modified Bessel function of the $5/3$ order, 
the characteristic synchrotron photon energy $\epsilon_c$ is
\begin{equation}
	\epsilon'_c = \frac{3 h e B'}{4\pi m_e c} \, \gamma^2 \, , \label{eq:charac}
\end{equation}
and $B'$ is the jet magnetic field intensity. In the case of a relativistic 
bulk motion of the emitting plasma, the observed energy of the synchrotron 
photon is $\epsilon = \delta \, \epsilon'$, and the spectral emissivity 
transforms as $j_{\rm syn}(\epsilon) = \delta^2 \, j'_{\rm syn}(\epsilon')$, 
where $\delta$ is the corresponding Doppler factor\footnote{These 
transformations, together with the transformation of the emitting 
volume $V = \delta \, V'$ as appropriate for a homogeneous moving 
source assumed here, give the energy flux transformation 
$[\epsilon S_{\epsilon}] = \delta^4 \, [\epsilon' S'_{\epsilon'}]$, 
since $[\epsilon S_{\epsilon}] \propto [\epsilon j_{\epsilon}] \times V$.}. 
In the following analysis we adopt the product of $B' \delta = 2$ 
(where $B'$ is expressed in units of Gauss) for the blazar emission 
zone in Mrk~421, which is the value claimed previously for this 
source by \citet{tad00} based on detailed variability and spectral 
studies, when the object showed a similar flux to that measured by us. 
The validity of adopting $B' \delta = 2$ (and an investigation of 
a range for values adopted for this parameter) is discussed 
later. Hence, the observed synchrotron flux at a given observed 
photon energy in our model depends solely on the parameters of the EED. 

\subsection{Spectral Fitting} 
\label{sect:fitting}

Our observations took place during a relatively low activity state 
of Mrk~421, characterized by only a mild variability: the observed 
$2-10$\,keV flux varied between $\simeq 2.7\times10^{-10}$ and 
$\simeq 4.1\times10^{-10}$\,erg\,s$^{-1}$\,cm~$^{-2}$ 
during the 3-day long {\it Suzaku} observations. 
In order to construct a truly simultaneous broad-band spectrum ranging 
from UV to hard X-ray frequencies, we have extracted the available data 
taken only during the overlapping time intervals for XRT, XIS and HXD/PIN 
instruments; this resulted in $0.6$\,ksec of the final net exposure. 
Nevertheless, even with such a short exposure, we were still able to 
obtain a very good quality broad-band X-ray spectrum of Mrk~421 extending 
from $0.5$ up to $50$\,keV. The averaged X-ray spectrum constructed by 
accumulating all of the detected photons instead would not be significantly 
different than this truly simultaneous spectrum because, as noted above, 
Mrk~421 was relatively steady during the observations, and 
no significant spectral changes have been noted in the total 
accumulated {\it Suzaku} dataset. The conventional fit to the X-ray spectrum 
with a single power-law function modified solely by the Galactic 
absorption describes the data poorly ($\chi_\nu^2=3.47$ with 56~dof), and 
indicates a significant curvature of the X-ray spectrum. A broken power-law 
function provides a better representation of the data ($\chi_\nu^2=1.06$ with 
54~dof), returning the low- and high-energy photon indices of 
$\Gamma_{\rm low}=2.33\pm 0.01$ and $\Gamma_{\rm high}=2.71\pm 0.04$, 
respectively, with the break photon energy of $\epsilon_{\rm brk}=2.82\pm 
0.17$\,keV. The corresponding $2-10$\,keV energy flux of Mrk~421 is then  
$S_{\rm 2-10\,keV} \simeq 2.83\times 10^{-10}$\,erg\,s$^{-1}$\,cm$^{-2}$,
again, similar to that reported by \citet{tad00}. 

\begin{figure}[tbp]
\epsscale{.80}
\begin{center}
	\plotone{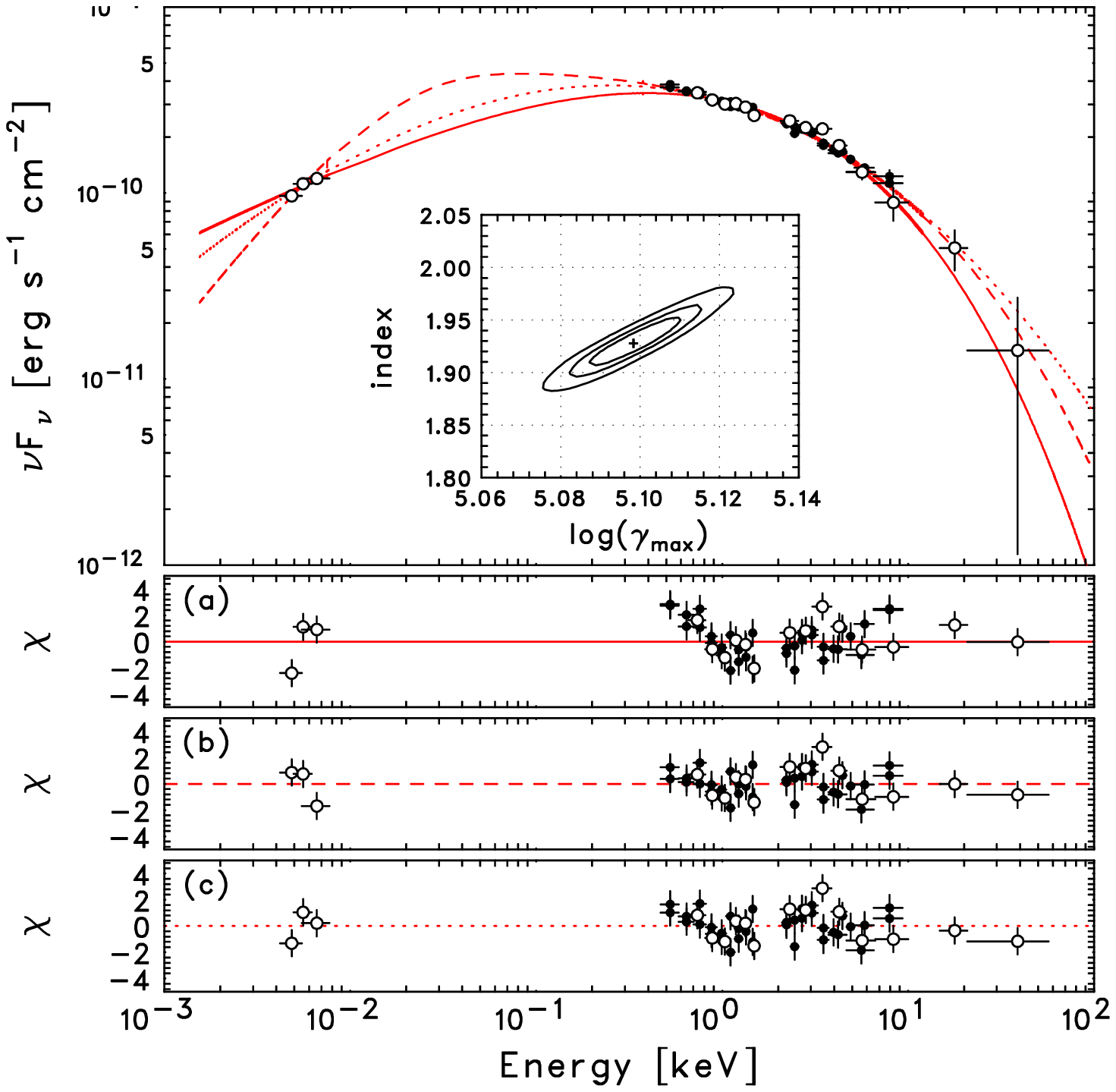}
\end{center}
\caption{The UV--to--X-ray spectral energy distribution of Mrk~421 fitted by the {\tt SYNCHROTRON} model. The top panel shows the analyzed data fitted with (a) a power-law EED with fixed $\gamma_{\rm min}=1$ (solid curve), (b) a power-law EED with $\gamma_{\rm min}$ set free (dashed curve), and (c) a modified ultra-relativistic Maxwellian EED (dotted curve). The bottom panels represent the residuals between the data and the models. The XIS data points are represented by filled circles in order to distinguish from the XRT and UVOT data.}
\label{fig:SED_keV}
\end{figure}

The constructed dataset is then analyzed using the {\tt SYNCHROTRON} model.  
The top panel of Figure~\ref{fig:SED_keV} presents the UV--to--X-ray spectral 
energy distribution of Mrk~421 fitted with three different representations
of the EED as described below, while the bottom panels show the residuals 
between the data and the models. The spectral fitting is performed jointly 
in the UVOT band and the $0.5-50$\,keV band;  here, we ignore the data 
within the $1.6-2.0$\,keV and the $10-15$\,keV photon energy ranges, 
because of the systematic calibration uncertainties \citep{serlemit06, 
fukazawa09}. The extinction and the absorption of the UV/X-ray photons in 
our Galaxy is taken into account by means of multiplicative constants 
based on the analysis by \citet[see Table~\ref{tab:param}]{pei1992}. Such 
a procedure was applied previously in the analysis of the spectra of 
Gamma-Ray Bursts \citep[e.g.][]{schady07}. An alternative analysis 
of the UV data following the procedure presented in \citet{kataoka08a} 
returns the fluxes consistent within a few percent with the ones obtained 
by our method. 

In the first representation of the fitted EED --- denoted below as the case 
``a'' ---  we consider the power-law (PL) model with fixed minimum electron 
Lorentz factor $\gamma_{\rm min}=1$. The results of the model fitting are 
summarized in Table~\ref{tab:param} and shown in Figure~\ref{fig:SED_keV}. 
The emerging parameters are $s=1.91^{+0.01}_{-0.02}$ and 
$\log(\gamma_{\rm max})=5.09^{+0.01}_{-0.01}$, with the quality of the fit 
$\chi_\nu^2=1.86$ (for $54$\,dof). The inset in Figure~\ref{fig:SED_keV} 
shows that these two parameters ($s$ and $\gamma_{\rm max}$) are weakly 
correlated with each other in a sense that the harder the electron 
distribution gets, the lower maximum energy of electrons is required 
to fit the data. However, the remaining residuals around $0.5$\,keV indicate 
that the case ``a'' may not be the best representation of the data. 
We emphasize again that here we probe directly the parameters of the 
EED, unlike the previous studies in which the observational data were 
fitted by some assumed parametric form of the \emph{emission} continuum 
\citep[either a power-law, a broken power-law, or a log-parabolic form; 
see e.g.,][]{tra09}. In addition, the EED parameters are determined 
robustly, because the UV data are taken into account in addition to 
the X-ray data. Without truly simultaneous UV data, the model fits 
would have a larger degeneracy. 

In the second representation of the fitted EED (hereafter case ``b'') 
we set free all the three parameters of the power-law model PL. The resulting 
values are then $s=2.77^{+0.13}_{-0.12}$, 
$\log(\gamma_{\rm min})=4.32^{+0.03}_{-0.03}$ and 
$\log(\gamma_{\rm max})=5.36^{+0.06}_{-0.05}$ with $\chi_\nu^2=1.12$ 
(for $53$\,dof). The quality of the fit is now significantly improved, 
as shown in Figure~\ref{fig:SED_keV}. Finally, in the third representation 
of EED denoted as the case ``c'', we consider the modified ultra-relativistic 
Maxwellian form (model UM). In this case, which represents the data well 
($\chi_\nu^2=1.20$ with $53$\,dof; see again Table~\ref{tab:param} and 
Figure~\ref{fig:SED_keV}), the model parameters are 
$\log(\gamma_{\rm eq})=1.41^{+0.06}_{-0.03}$ and $1/a=5.14^{+0.06}_{-0.09}$.

%%%%%%%%%%%%%%%%%%%%%%%%%%%%%%%%%%%%%%%%%%%%
\begin{table}[tbp]
\small
\caption{The fitting results with the {\tt SYNCHROTRON} model\tablenotemark{a}}
\begin{center}
\begin{tabular}{ccccccc}
\hline
model & case &  & model parameters &  & $2-10$\,keV flux & $\chi^2_\nu$ (dof) \\
 & & $\log(\gamma_{\rm min})$ & $\log(\gamma_{\rm max/eq})$ & $s$ or $1/a$ & $S_{\rm 2-10\,keV}$ [erg\,cm$^{-2}$\,s$^{-1}$] & \\
\hline
PL & ``a'' & $0.0$ (fixed)  & $5.09\pm 0.01$         & $s=1.91^{+0.01}_{-0.02}$   & $2.61 \times 10^{-10}$ & 1.86 (54) \\
PL & ``b'' & $4.32\pm 0.03$ & $5.36^{+0.06}_{-0.05}$ & $s=2.77^{+0.13}_{-0.12}$   & $2.65 \times 10^{-10}$ & 1.12 (53) \\
UM & ``c'' & $0.0$ (fixed)  & $1.41^{+0.06}_{-0.03}$ & $1/a=5.14^{+0.06}_{-0.09}$ & $2.65 \times 10^{-10}$ & 1.20 (54) \\
\hline
\end{tabular}
\end{center}
\label{tab:param}
\tablenotetext{a}{Obtained with the multiplicative constants {\tt const}=UVOT:XRT:XIS:PIN=1.00:1.07:1.00:1.13, E(B$-$V)=0.015, R$_{\rm V}$=3.1, 
and z=0.0 for {\tt zdust} \citep{schlegel98}, and N$_{\rm H}$=0.0161 for 
{\tt wabs} \citep{lockman95}.}
\end{table}
%%%%%%%%%%%%%%%%%%%%%%%%%%%%%%%%%%%%%%%%%%%%

\subsection{Related Uncertainties}
\label{sect:uncertainties}

\begin{figure}[tbp]
\epsscale{.80}
\begin{center}
	\plotone{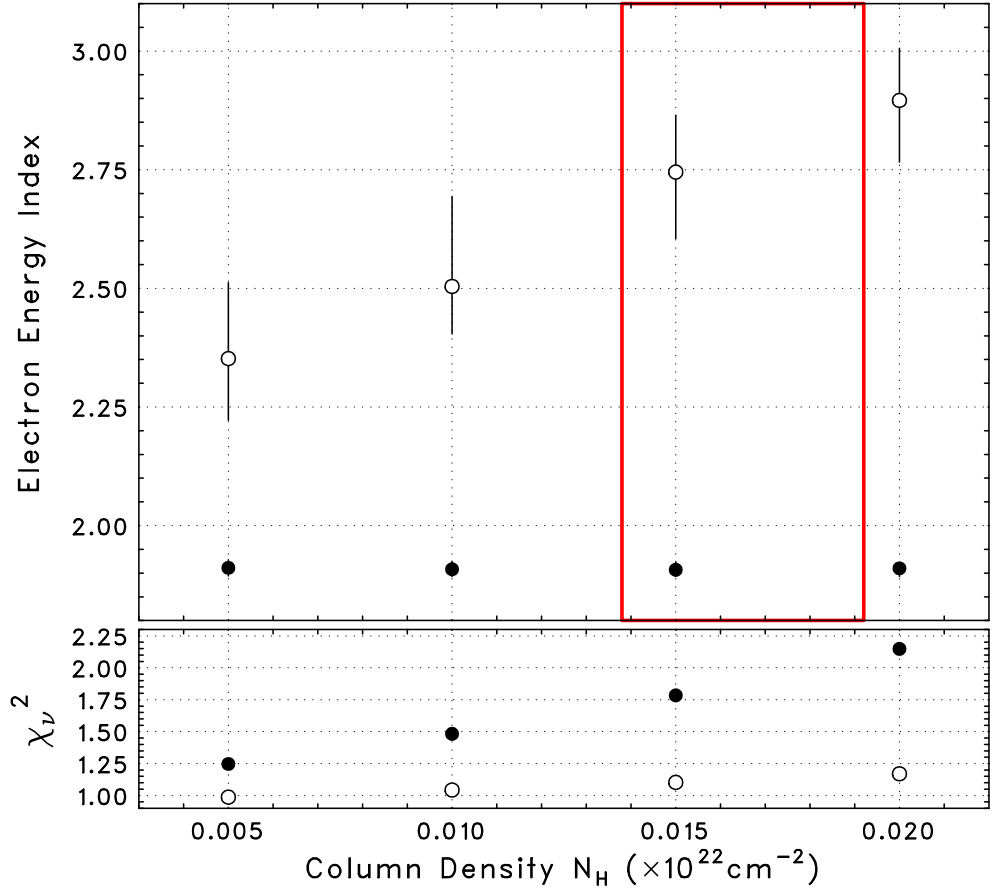}
\end{center}
\caption{The dependence of the electron power-law slope (cases ``a'' and ``b'' represented by filled and open circles, respectively) on the Galactic column density, $N_{\rm H}$ (top panel). The extinction parameter E(B-V) is scaled linearly with $N_{\rm H}$ (see text). The allowed range for $N_{\rm H}$ inferred from the historical H{\sc I} observations is indicated by a red rectangle. The corresponding reduced $\chi^2$ values are plotted in the bottom panel.}
\label{fig:nH}
\end{figure}

The model parameters emerging from the analysis described above may 
depend to some level on the assumed value of the Galactic column density 
in the direction of Mrk~421, which affects the low energy X-ray data. 
Therefore, it is necessary to investigate how robust are the presented 
model fits by varying $N_{\rm H}$ within the allowed range and repeating 
the analysis. In the case of the PL models (cases ``a'' and ``b''), 
Figure~\ref{fig:nH} shows the resulting dependence of the electron 
energy index $s$ on the level of the Galactic absorption, with the 
extinction parameter E(B--V) varied accordingly (assuming linear scaling 
with $N_{\rm H}$). In the figure, the red box indicates the allowed range 
for $N_{\rm H}$ estimated from the historical H{\sc I} observations 
\citep{dickey90, lockman95, kalberla05}. As is shown there, in the case ``a'' 
(filled circles in the plot) the slope of the EED is almost independent 
on the particular value of the absorption/extinction coefficients, but 
the acceptance of the fit gets significantly worse as $N_{\rm H}$ becomes 
larger. This behavior is easy to understand, since with the fixed minimum electron 
energy $\gamma_{\rm min} = 1$ the remaining two parameters of the power-law 
model are determined by the broad-band UV--to--X-ray emission continuum, 
almost irrespectively on the low-energy segment of the X-ray dataset 
($<1.0$\,keV) which might be most affected by increased $N_{\rm H}$ and which, 
as such, determines only the quality of the fit. The same is true for the UM 
model (case ``c'', not included in Figure~\ref{fig:nH}); namely, changes in 
the hydrogen column density (and, correspondingly, the extinction level) 
do not change significantly the emerging parameters of this model, 
but only affect the quality of the fit. We note that with the assumed 
linear scaling of E(B-V) with $N_{\rm H}$ the fluxes in the UVOT band vary 
much less substantially (i.e., within 10\%) than the soft X-ray flux.

On the other hand, in the case of the PL model with all three parameters set 
free (case ``b'', open circles in Figure~\ref{fig:nH}), the emerging electron 
energy index correlates with $N_{\rm H}$, while the quality of the fit remains 
very good. As a result, within the allowed range of the Galactic absorption 
the electron index in the case ``c'' reads roughly as $s \simeq 2.5-3.0$ 
(compare it with Table~\ref{tab:param}). Again, the observed dependence 
is not difficult to understand. This is because in the case ``b'' the 
minimum electron Lorentz factor is so large that the observed synchrotron 
emission in the UV frequency range is produced almost entirely by the 
lowest-energy electrons. In particular, with the assumed $B' \delta = 2$ 
the observed characteristic frequency of the synchrotron photons emitted 
by the electrons with $\gamma_{\rm min} = 2 \times 10^{4}$ is 
$\epsilon_c \simeq 15$\,eV. In other words, the slope of the electron 
energy energy distribution in the case ``b'' is determined not predominantly 
by the broad-band UV--to--X-ray emission, as in the cases ``a'' and ``c'', 
but solely by the X-ray continuum. As such, the parameter $s$ in the 
case ``b'' is sensitive to the low-energy X-ray photons, and therefore 
to the particular value of $N_{\rm H}$.

Another concern arises from the fact that in our analysis the product 
of the magnetic field intensity and the jet Doppler factor has been 
fixed at the particular value $B' \delta = 2$. There have been numerous 
efforts to determine uniquely the physical parameters of blazar emission 
region in TeV-detected BL Lacs, including $B'$ and $\delta$, in addition 
to the aforementioned work by \citet{tad00}. For example, \citet{tave98}, 
who attempted to utilize the multi-wavelength data self-consistently, 
claimed $\delta=25$ and $B'=0.2$\,G for the flaring state of Mrk~421 
\citep[see][]{tad96}. This would lead to $B' \delta = 5$. Similarly, 
\citet{alb07a} obtained $B'=0.2$\,G and $\delta=15$, resulting in 
$B' \delta = 3$. Although those values are slightly different from the 
one assumed in the our paper, they are of the same order of magnitude.  
In order to investigate in detail how the related uncertainty regarding 
the product $B' \delta$ affects the presented results, we repeated the 
analysis for $B' \delta=1$, 5, and 10. We found that within the range 
$B' \delta = 1-10$, the resulting curvature of the EED (parameters $s$ or $a$) 
does not significantly change when compared to 
$B' \delta=2$, but only that the critical electron energies 
($\gamma_{\rm min}$, $\gamma_{\rm max}$ and $\gamma_{\rm eq}$) scale as 
$\propto 1/\sqrt{B' \, \delta}$, as expected (see equation~\ref{eq:charac}). 
Therefore, we can safely conclude that the uncertainty regarding $B' \delta$ 
does not affect significantly the model results. 

\section{Discussion:  Insights into the Energy Distribution of Radiating Particles} 
\label{sect:discussion}

In the previous sections we showed that the UV--to--X-ray synchrotron 
continuum of Mrk~421 observed in December 2008 may be fitted almost 
equally well by three different representations of the EED, each well 
motivated on theoretical grounds, yet each implying different physical 
conditions and different particle acceleration processes operating within the 
blazar emission zone of Mrk~421. Here we discuss in this context and in more 
detail the results obtained by us. In the following, we assume that the 
inferred shapes of the EED may be identified with the steady-state spectra 
formed within a homogeneous acceleration/emission region. It is not clear (or 
it is even questionable), however, whether this particular assumption --- even 
though being widely applied in blazar modeling --- is a good approximation 
of a realistic situation, and we comment on this issue in \S\,\ref{sect:Steady} below.

\subsection{Ultra-relativistic Maxwellian} 
\label{sect:UltraMaxwell}

We begin with the case ``c'' analyzed above, namely with the ultra-relativistic 
Maxwellian representation of the energy distribution of the radiating 
electrons.  In general, it was shown previously by various authors that 
the Maxwellian spectra $\propto \gamma^2 \, \exp[-\gamma/\gamma_{\rm eq}]$ 
are expected to form in the case of an efficient continuous acceleration 
of relativistic particles undergoing radiative (synchrotron-type) losses 
obeying $\tau'_{\rm rad} \propto \gamma^{-1}$ and a very inefficient escape 
from the acceleration/emission region \citep{sch84,aha86,par95}.  
In particular, turbulent (``2nd-order'', or ``stochastic'') acceleration
of the Fermi-type --- characterized by 
the acceleration timescale independent on the particle energy, 
$\tau'_{\rm acc} \propto const$ --- was the most commonly considered 
example of the continuous energization of ultra-relativistic particles, 
with the equilibrium energy $\gamma_{\rm eq}$ defined via the 
condition $\tau'_{\rm acc} = \tau'_{\rm rad}(\gamma_{\rm eq})$.  
This has been generalized by \citet{sta08}, who showed that for 
the power-law scalings $\tau'_{\rm acc} \propto \gamma^p$ and $\tau'_{\rm rad} 
\propto \gamma^q$, the steady-state solution to the appropriate equation 
governing energy evolution of the radiating particles (assuming, again, 
very inefficient particle escape from the system) is of the modified 
ultra-relativistic Maxwellian form 
$\propto \gamma^2 \, \exp[- \, (1/a) \,\, (\gamma/\gamma_{\rm eq})^a]$, 
where $a = p-q$.

Our analysis indicates that in order to fit the UV--to--X-ray spectrum 
of Mrk~421 with the modified Maxwellian form of the EED, the equilibrium 
energy has to be very small, $\gamma_{\rm eq} \simeq 30$, and the spectral 
curvature has to be very smooth, $a \simeq 0.2$. Even though it is formally 
possible to relate these requirement to a set of particular physical conditions 
within the emission zone of Mrk~421, the derived parameters appear 
unrealistic. For example, anticipating 
the dominant inverse-Compton cooling to take place in the Klein-Nishina 
(KN) regime, i.e. $q\simeq0.5$, one would have to assume $p \simeq 0.7$ 
in order to obtain $a \simeq 0.2$. This requirement, interestingly, would 
then be consistent with the stochastic acceleration of the radiating 
particles by the turbulence characterized by the energy spectrum 
similar to the Kreichnan one\footnote{For the energy spectrum of the 
turbulence $\mathcal{W}(k) \propto k^{-\sigma}$, where $k$ is the 
wavevector of the turbulent modes, stochastic acceleration timescale reads 
as $\tau'_{\rm acc} \propto \gamma^{2-\sigma}$. Hence, for the Kreichnan 
spectrum $\sigma = 3/2$ one obtains $\tau'_{\rm acc} \propto \gamma^{1/2}$ 
\citep[see][and references therein]{sta08}.}. However, the anticipation that 
relatively low-energy electrons within the blazar emission zone of Mrk~421 
cool predominantly via the inverse-Compton emission in the KN regime rather 
than via synchrotron emission is highly questionable. For the case where the 
synchrotron emission dominates ($q=-1$), on the other hand, no commonly acceptable 
value of $p$ can be found to satisfy $a \simeq 0.2$. Hence, we consider 
the case of the UV--to--X-ray spectrum of Mrk~421 fitted by the modified 
Maxwellian form of EED as only a formal possibility only, involving somehow 
unphysical (or, at least, ``non-standard'') conditions.

\subsection{Shock-originated Power-law} 
\label{sect:PowLawElectron}

Next we discuss the power-law models for the EED, focusing first 
on the case ``b''. In this case, the energy spectrum of radiating 
electrons is characterized by a steep slope $s \simeq 2.77$ within a very 
(uncomfortably?) narrow energy range $2 \times 10^4 \lesssim \gamma \lesssim 2 \times 10^5$ 
(preceded by a sharp low-energy cut-off or eventually a very hard low-energy 
tail). At the first glance, such a representation of the EED may seem as 
artificial as the one discussed in the previous section, but we note here that 
it is in fact very similar to that invoked by \citet{kat06a} for the 
case of a distant ($z=0.186$) BL Lac object 1ES~1101-232 surprisingly 
detected at TeV energies, where the absorption of its TeV emission 
due to pair production on the extragalactic diffuse background light 
was expected to provide large opacity at TeV energies \citep[see also in this
context][for the case of $z=0.14$ BL Lac 1ES\,0229+200]{tav09}. 
In addition, the high minimum electron energy and the steep electron 
spectral index (significantly different from the ``canonical shock 
spectrum'' $s = 2.0$) might in principle be both reconciled with the diffusive 
(``1st-order-Fermi'') acceleration of the radiating electrons at a 
relativistic, perpendicular\footnote{``Perpendicular/oblique'' shocks 
are those with magnetic field lines oriented at right/large angles to 
the shock normal. Furthermore, if the projection of the upstream fluid 
velocity on the upstream magnetic field direction is smaller/larger than the 
light speed, the respective shocks are called ``subluminal/superluminal''.}, 
and proton-mediated shock. That is because under such conditions the inertia 
of protons carrying bulk of the jet energy (and thus shaping the shock 
structure) define the critical electron energy 
$\gamma_{\rm cr} \simeq \Gamma_{\rm sh} \, m_p/m_e \simeq 2 
\times 10^4 \, (\Gamma_{\rm sh} / 10)$ (where $\Gamma_{\rm sh}$ 
is the bulk Lorentz factor of the shock in the upstream rest frame), 
below which no 1st-order-Fermi acceleration is in principle allowed \citep{beg90,hos92}.  
It is thus interesting that the minimum electron energy obtained in the case 
``b'' is in agreement with the expected value of $\gamma_{\rm cr}$, but
only in the case of ultrarelativistic shock velocity, $\Gamma_{\rm sh} 
\simeq 10$. In addition, the steep power-law slope of the electrons with energies 
higher than $\gamma_{\rm cr}$ is in agreement with those 
claimed for the diffusive acceleration at superluminal shocks 
\citep{nie04,sir09}, assuring self-consistency of the interpretation.

Even though similar arguments as above (based on the spectral forms of the 
EED derived from the broad-band modeling) were given previously in favor of 
a dominant role of perpendicular, proton-mediated shocks in AGN jets 
\citep[e.g.,][]{sta07,sik09}, the concern regarding the presented interpretation 
of the dataset analyzed in this paper is that large $\gamma_{\rm min}$ obtained 
in the case ``b'' may be an artifact of the analysis forcing the broad-band 
UV--to--X-ray data to be fitted by an EED of single power-law form. Indeed, 
as already noted in \S\,\ref{sect:uncertainties}, 
with the assumed $B' \delta = 2$ the observed characteristic frequency of 
the synchrotron photons emitted by the electrons with 
$\gamma_{\rm min} = 2 \times 10^{4}$ is $\epsilon_c \simeq 15$\,eV, i.e. 
just above the UV frequency range. Thus, the obtained value for the minimum 
electron Lorentz factor in the case ``b'' may not reveal the nature of 
particle acceleration within the blazar emission zone of Mrk~421, but only 
the fact that the analyzed UV and X-ray data cannot be fitted together by 
a single power-law form of the EED. 

Besides the aforementioned caveat, the particular form of the EED discussed 
here would require a lot of fine tuning to ensure $1 \ll \gamma_{\rm min} 
\lesssim \gamma_{\rm max}$ within the emission region, or in 
other words to prevent formation of a low-energy tail in the electron distribution
(energies $\gamma < \gamma_{\rm min}$) due to the radiative cooling of
ultrarelativistic particles injected from the shock to the emission site. Note 
in this context that with a postulated efficient re-acceleration processes taking 
place downstream of the shock, which could potentially remove the 
cooling problem, one should not expect the radiating electrons to maintain 
their injection spectrum, but rather to form a continuum resembling 
modified Maxwellian discussed in \S\,\ref{sect:UltraMaxwell}. Yet the possibility 
for preserving the ``shock-like'' EED with high $\gamma_{\rm min}$ far from the 
injection site cannot be rejected robustly. Clearly, such a controversy could in 
principle be resolved with the simultaneous {\it Fermi} observations, because the 
position of $\gamma_{\rm min}$ is the major factor determining the 
slope of the $\gamma$-ray continuum within the MeV--GeV photon energy 
range. In other words, {\it Fermi}/LAT data, if simultaneous with the 
UV--to--X-ray data analyzed with the {\tt SYNCHROTRON} model, could 
help to discriminate between the case ``b'' and other analyzed cases, 
i.e. between the high or low minimum electron Lorentz factor.

\subsection{Flat Power-law as a Cooled Maxwellian?} 
\label{sect:CooledMaxwell}

Finally, we consider the power-law model, case ``a'', in which the minimum electron 
Lorentz factor is fixed at $\gamma_{\rm min} = 1$, and the emerging power-law slope 
of the EED is $s \simeq 1.93$. We note that because only the UV--to--X-ray data are 
fitted in our analysis, any other fixed value for $\gamma_{\rm min}$ within the 
range $1-10^3$ would not affect the determination of the parameter $s$, since 
the observed characteristic frequency of the electrons with Lorentz factors 
$\gamma < 10^3$ is expected to be located below the UVOT frequency range 
(with the assumed $B' \delta = 2$). Thus the physical case to discuss is 
a flat ($s \simeq 1.91$) electron power-law between the energies 
$\gamma_{\rm min} \leq 10^3$ and $\gamma_{\rm max} \simeq 10^5$. 
There may be two interpretation for such. One is the diffusive shock 
acceleration at a (mildly) relativistic subluminal shock (without necessarily 
significant proton content);  such a configuration results in a very flat particle 
spectra \citep[$s < 2.0$;][]{kir89,ost91,bed96}. 
In a framework of this interpretation, the entire observed synchrotron continuum 
of Mrk~421 would then be produced in a slow-cooling regime.

The other possible physical interpretation for the case ``a'' discussed in 
this paper may be a radiatively re-processed (cooled) Maxwellian electron 
distribution. As is well known, with a steep ($s \geq 2.0$) power-law EED 
continuously injected into the emission region, synchrotron-type radiative 
cooling ($\tau'_{\rm rad} \propto \gamma^{-1}$) results in further steepening 
of the injected spectrum (by $\Delta s = 1.0$) above some critical (break) energy. 
On the other hand, in the case of a very flat, high-energy EED injected to the 
emission region, radiative cooling leads to formation of a low-energy power-law 
tail $\propto \gamma^{-2}$ instead. For example, a radiatively re-processed ultra 
relativistic Maxwellian $\gamma^2 \, \exp[-\gamma/\gamma_{\rm eq}]$ would give 
a familiar $\gamma^{-2} \, \exp[-\gamma/\gamma_{\rm eq}]$ form of the EED.  
Moreover, as pointed out by \citet{gie07}, additional non-negligible 
inverse-Compton/KN losses may significantly flatten such a power-law tail, 
leading to a formation of cooled power-law spectra with $s < 2.0$.  
We believe that this mechanism is a viable option for the interpretation 
of the synchrotron emission of Mrk~421. And indeed, \citet{gie07} applied 
the discussed scenario in the particular context of Mrk~421, showing that 
a single power-law with $s=1.7$ --- formed as a cooled Maxwellian --- may 
account for the broad-band emission of this blazar observed in March 2001. 
The value of the electron slope claimed by \citet{gie07} is lower than the 
one found in our analysis but, interestingly, the average $2-10$\,keV flux 
of Mrk~421 in March 2001 
($S_{\rm 2-10\,keV} \simeq 13.4 \times 10^{-10}$\,erg\,cm$^{-2}$\,s$^{-1}$), 
was about 5 times higher than during the exposure analyzed here. Such a 
behavior may be reconciled with the interpretation above if one assumes 
increasingly pronounced KN effects in cooling of the radiating electrons 
as the synchrotron flux increases. 

The above interpretation requires however a strong cooling regime for all 
the electrons down to energies of $\gamma \simeq 10^3$. This, in turn,
would imply relatively large emission region size $R$: by simply equating
the dynamical timescale $\tau'_{\rm dyn} \simeq R/c$ with the cooling
(predominantly synchrotron by assumption) timescale $\tau'_{\rm rad} \simeq 
\tau'_{\rm syn} \simeq 6 \pi m_e c/ \sigma_{\rm T} \gamma B^2$, 
one may find very roughly
\begin{equation}
R \sim 0.03 \, \left({\gamma \over 10^3}\right)^{-1} \, \left({B' \over 0.5\,{\rm G}}\right)^{-2} \, {\rm pc} \, .
\end{equation}
Since the corresponding variability timescale $\tau_{\rm var} \simeq R/ c \, \delta$
would then be of the order of days--weeks (for $\gamma = 10^3$, $B' \lesssim 1$\,G, 
and the jet Doppler factor $\delta \lesssim 10$),
any intraday flux variations could be hardly accounted by the discussed model. 

The interesting scenario outlined above can be easily and directly tested 
with the simultaneous UV, X-ray, and $\gamma$-ray data (obtained with the 
{\it Fermi} satellite and modern IACTs). That is because such a broad-band 
dataset would allow one to determine the ratio of the synchrotron and 
inverse-Compton cooling rates (or, more precisely, the relative contribution 
of the inverse-Compton/KN cooling), together with the parameters of the 
underlying EED using the {\tt SYNCHROTRON} model. Note in this context that 
in the first interpretation of the case ``a'' --- the one involving shock 
acceleration --- no obvious correlation between the cooling rates ratio and 
the electron spectral index is expected, while in the second interpretation 
--- the one involving radiatively re-processed Maxwellian component --- a clear 
correlation between these two quantities is implied. We want to point out that 
there exists Fermi and MAGIC data during the 3-day long {\it Suzaku} observations. 
The analysis of these data set, as well as the physical interpretation of the 
overall SED in the context of the above presented model will be reported on a 
forthcoming publication.

\subsection{Steady Component?} 
\label{sect:Steady}

In the discussion above we have assumed that all the possible 
representations of the EED emerging from the analysis correspond 
to the steady-state spectra formed within a homogeneous 
acceleration/emission region of Mrk~421 jet. As already emphasized, 
this assumption may not be correct. And indeed, if the blazar emission 
zone is highly inhomogeneous, then the inferred electron spectra could instead 
be due to a superposition of multiple distributions formed within 
different sub-volumes of a system. In particular, the Maxwellian 
form discussed above --- with the ``problematic'' parameter $a \simeq 0.2$ --- 
may reflect some complex superposition of different Maxwellians with 
``physically acceptable'' curvatures $a > 0.5$ or, alternatively, some 
multiply-broken power-law electron continuum. As such, it may well 
correspond to the steady emission component claimed for Mrk~421 by 
\citet[see section 5 therein]{ushio2009}. Interestingly, the X-ray 
flux of this component and its spectral curvature \citep[which we 
determined for the purpose of this paper by fitting the UM form with 
the {\tt SYNCHROTRON} model to the dataset analyzed in][]{ushio2009} 
are both quite similar to those discussed here.

\section{Conclusions} 
\label{sect:conclusions}

In this paper we have investigated which form of the electron energy 
distribution describes the best the UV--to--X-ray spectrum of Mrk~421 
in its relatively low state (December 2008; the observed $2-10$\,keV 
flux $S_{\rm 2-10\,keV} \simeq 2.8 \times 10^{-10}$\,erg\,cm$^{-2}$\,s$^{-1}$). 
In our studies we have applied the forward-fitting {\tt SYNCHROTRON} model 
implemented in {\tt XSPEC} \citep{ushio2009}; that is, the analyzed dataset 
has been fitted directly with a synchrotron emission originating from a given 
(assumed) shape of the electron distribution, rather than by an assumed parametric 
form of the non-thermal emission continuum as typically presented in the 
literature. We have considered two different general representations of the 
electron spectrum, both motivated by the particle acceleration models discussed 
in the literature, namely a power-law form within some finite energy range 
modulated by an exponential cutoff at the maximum energy, and a modified 
ultra-relativistic Maxwellian. We have found that the latter one, which 
would correspond to the continuous (turbulent) acceleration of the radiating 
particles within the emission region \emph{with no particle escape}, even 
though formally allowed (i.e. giving a reasonable fit to the data), implies rather 
artificial, and thus unlikely physical conditions within the blazar zone of 
Mrk~421. On the other hand, a power-law representation of the electron energy 
spectrum fits the data equally well. In this case, two possible sets of the 
model parameters have been found: (i) a flat-spectrum one $\propto \gamma^{-1.91}$ 
with low electron minimum energy $\gamma_{\rm min} < 10^3$, and (ii) a 
steep-spectrum one $\propto \gamma^{-2.77}$ with high electron minimum energy 
$\gamma_{\rm min} \simeq 2 \times 10^4$. We have discussed different 
interpretations for both, involving diffusive acceleration of the radiating 
electrons at relativistic, sub- or superluminal shocks, as well as radiatively 
re-processed electron distribution of the initial (injected) Maxwellian form.

The presented technique in analyzing broad-band data of blazar sources constitutes 
a very promising method for investigating particle acceleration processes 
operating in relativistic plasma. In the particular case of the BL Lac object 
Mrk~421, it already allowed us to limit the number of possible scenarios to be 
considered in this context. Still, several different options remain, which could be 
investigated in detail using the {\tt SYNCHROTRON} model with new broad-band, 
good-quality UV--to--X-ray data covering long time intervals and different activity 
states in this and other similar sources. 
In this paper we focused on the simultaneous UV to X-ray Mrk~421 data from {\it Swift} and {\it Suzaku} observations in December 2008, and we argued that gamma-ray data simultaneous  to the UV-X-ray data would add further constrains to the various possible model/physical interpretations. We want to point out that there exists Fermi/MAGIC observations contemporaneous to the {\it Swift}/{\it Suzaku} observations during December 2008. The analysis of this data set, as well as the physical interpretation of the overall SED in the context of the above presented model will be reported on a forthcoming publication.

\acknowledgments
{\L}.S. and M.O. are grateful 
for the support from Polish MNiSW through the grant N-N203-380336. GM acknowledges 
support via NASA grants NNX08AX77G and NNX08AZ89G.

%%%%%%%%%%%%%%%%%%%%%%%%%%%%%%%%%%%%%%%%%%%
%%%%%%%%%%%%%%%%%%%%%%%%%%%%%%%%%%%%%%%%%%%

%%%%%%%%%%%%%%%%%%%%%%%%%%%%%%%%%%%%%%%%%%%
%%%%%%%%%%%%%%%%%%%%%%%%%%%%%%%%%%%%%%%%%%%

\end{document}